\documentclass[%
 preprint, 
 amsmath,amssymb,
 aps, physrev,
]{revtex4-2}
\usepackage{physics}
\usepackage{tikz}
\usepackage{subcaption}
\usepackage{verbatim}
\usepackage{graphicx}
\usepackage{dcolumn}
\usepackage{bm}
\usepackage{hyperref}

\usepackage{lipsum}
\usepackage{xcolor}
\usepackage{mathrsfs}

\begin{document}


\title{\textbf{
Horizon brightened acceleration radiation entropy in causal diamond geometry: A near-horizon perspective} 
}%

\author{Nada Eissa,$^{1}$ Carlos R. Ordóñez,$^{1}$ and Gustavo Valdivia-Mera$^{1}$}

\affiliation{\vspace{5mm}
$^{1}$Department of Physics, University of Houston, Houston, Texas  77204-5005, USA\vspace{5mm}}
\date{\today}\vspace{5mm}
\begin{abstract}
In this article, we extend the horizon brightened acceleration radiation (HBAR) framework, originally introduced by Marlan Scully \textit{et al.} in Proceedings of the National Academy of Sciences \textbf{115}, 8131 (2018), to the causal diamond (CD) spacetime. We study a cloud of two-level atoms, injected at random times in the asymptotic past of the CD, freely falling toward its causal horizon and emitting scalar radiation via a weak dipole coupling to a quantum field. In the near-horizon region, an emergent conformal symmetry—captured by conformal quantum mechanics (CQM)—governs the field dynamics and allows analytic control of the emission process. We find that the radiation spectrum is thermal, with temperature $T_D = 1/(\pi\alpha)$, and that the associated von Neumann entropy flux reproduces the entropy production of the radiation field. These results demonstrate that the causal horizons of the CD spacetime effectively act as a \emph{topological thermal reservoir}, with thermal properties arising entirely from the global causal structure rather than from underlying microscopic degrees of freedom, highlighting that the validity of the HBAR framework is fundamentally tied to the existence of causal horizons, independent of the presence of a black hole.
\end{abstract}

\maketitle
\newpage
\tableofcontents
\newpage
\section{Introduction}

In the context of  horizon brightened acceleration radiation (HBAR) entropy, as proposed by Marlan Scully \textit{et al.} in Ref.~\cite{scully2018quantum}, it has been argued that the nature of the radiation emitted by a cloud of atoms—randomly injected from the asymptotic past and following free-falling (geodesic) trajectories toward the event horizon of a black hole—is a consequence of the relative acceleration between the field modes, defined on the black hole geometry with a well-defined Boulware vacuum state \cite{boulware1975quantum}, and the locally inertial motion of the atoms. The origin of this radiation lies in the weak dipole coupling between the atoms, modeled as two-level systems, and the quantum field \cite{PhysRevLett.91.243004}. A characteristic process of this interaction is the excitation of an atom accompanied by the emission of a scalar quantum, which can either reach future timelike infinity—thereby contributing to the observable radiation spectrum—or be reabsorbed by another atom within the cloud before being emitted again.\\

Due to the weak nature of the atom-field coupling, this process can be treated perturbatively, allowing for the computation of the scalar quanta emission probability. Remarkably, this emission rate is found to follow a Planckian distribution with an effective temperature given by $T = \kappa/(2\pi)$, where $\kappa$ is the surface gravity of the black hole. This temperature precisely reproduces the one derived by Hawking in his foundational analysis of black hole thermodynamics \cite{hawking1974black, hawking1975particle}. Furthermore, using tools from quantum laser theory \cite{scully1966quantum}, the density matrix of the radiation field can be characterized, and the von Neumann entropy then allows for the evaluation of the entropy flux $\dot{S}$, which not only exhibits thermodynamic behavior but also recovers the original results of Hawking and Bekenstein \cite{bekenstein2020black, bekenstein1973black}, including the area law.\\

At this point it is also important to highlight that, within the theoretical framework of black hole thermodynamics, the dynamics of a real scalar field in the near-horizon geometry of a black hole reduces to that of a one-dimensional conformal quantum mechanical system (CQM). In this limit an emergent conformal symmetry arises, which manifests in the behavior of the field modes—referred to as CQM modes—characterized by scale invariance within the domain of validity of the near-horizon approximation \cite{camblong2020near}. In this regime the relevant scale is determined by the conformal parameter $\Theta$, which depends on the mode frequency and the surface gravity. The CQM modes thus encode the essential information required to characterize black hole thermodynamics, as demonstrated in Ref.~\cite{camblong2005black}.\\

Consequently, the HBAR framework finds a natural extension in the near-horizon geometry, providing a powerful tool for analyzing more complex spacetime geometries, especially in situations where an exact geodesic analysis is intractable. The conformal symmetry that emerges in this regime allows for analytic control over the emission and absorption spectra of scalar quanta associated with acceleration radiation. This approach has proven fruitful in subsequent analyses of HBAR entropy for a variety of geometries, including Schwarzschild black holes \cite{camblong2020near}, Kerr black holes \cite{azizi2021acceleration, azizi1, azizi2}, quantum-corrected black hole geometries \cite{sen2022equivalence,jana2024atom,jana2025inverse}, along null geodesics \cite{chakraborty2019detector}, braneworld black holes \cite{das2024horizon}, and analyses involving pointlike and finite-size detectors \cite{das2025derivative}.\\

In the present work we analyze the proposal of HBAR entropy within the most fundamental causal structure in Minkowski spacetime: the causal diamond of size $2\alpha$. To this end, we examine the spacetime perceived by an observer with finite lifetime, constrained to this region—namely, the causal diamond spacetime. This setup demonstrates that the HBAR formalism can be generalized beyond black hole spacetimes, applying instead to more fundamental entities characterized by causal disconnection through horizons. The structure of the paper is as follows: In Sec.~II, we introduce the geometry of the causal diamond spacetime and the coordinate transformation that characterizes its metric. We then analyze the boundary of this spacetime, which contains both asymptotic regions and causal horizons. Next, we compute the effective surface gravity associated with its causal horizons from the perspective of the extended Minkowski spacetime, followed by an analysis of the geodesic structure within the near-horizon approximation. In Sec.~III, we explore the emergent symmetry in the near-horizon region, identified with conformal quantum mechanics, and characterize the corresponding scalar field modes. In Sec.~IV, we apply the HBAR framework in detail, explaining the interaction between the randomly injected atomic cloud and the field, computing the emission and absorption rates of scalar quanta, and using the master equation approach for the radiation field density matrix to justify the thermal character of the spectrum. Finally, in Sec.~V, we use the von Neumann entropy to derive a thermodynamic characterization of HBAR entropy. We conclude with a discussion and potential future directions.

\section{Causal diamond geometry}
\subsection{Coordinate transformation and metric tensor}\label{cmoftrw}
\begin{figure}[h]
 \centering
   \begin{subfigure}{0.4\textwidth}
    \includegraphics[width=\linewidth]{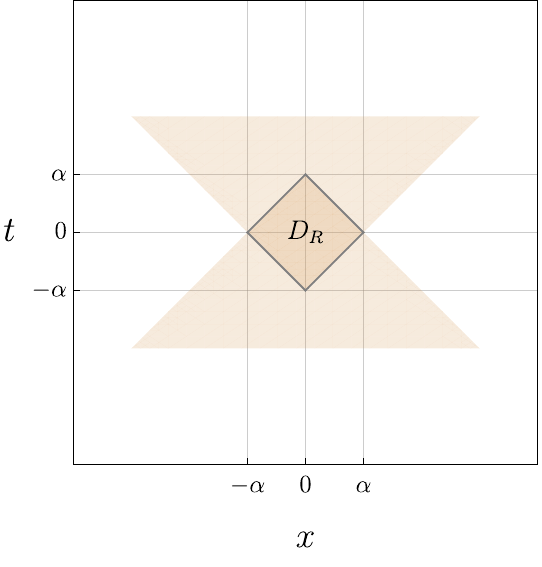}
     \caption{}\label{cfcp}
   \end{subfigure}
   \hspace{0.5cm}
   \begin{subfigure}{0.385\textwidth}
     \centering
     \includegraphics[width=\linewidth]{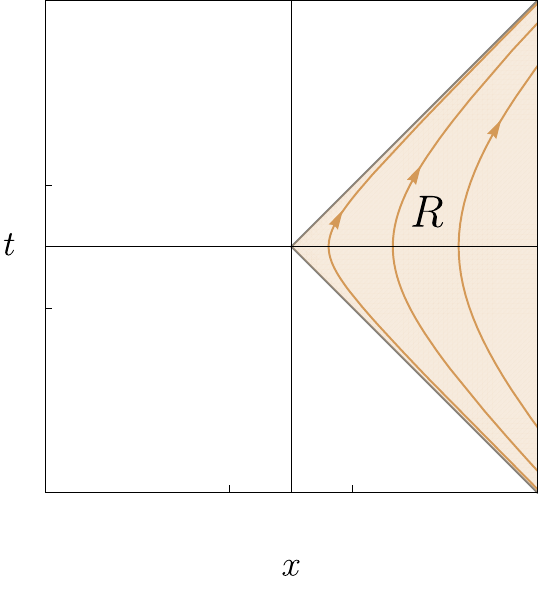}
     \caption{}\label{rrw}
   \end{subfigure}
  \caption{(a) The causal diamond $D_R$ defined as the intersection of a future and a past light cone. (b) Uniformly accelerated trajectories in the right Rindler wedge.}
\end{figure}

A causal diamond of length $2\alpha$ in Minkowski spacetime is defined by the region $D_R:=\qty{(t,x):\abs{t} + \abs{x} \leq \alpha}$. This region corresponds to the causal domain of an observer with a finite lifetime, bounded by the intersection of the future light cone of the birth event at $(t,x) = (-\alpha, 0)$ and the past light cone of the death event at $(t,x) = (\alpha, 0)$, as illustrated in Fig.~\ref{cfcp}. Therefore, the boundary of the causal diamond precisely contains the causal horizon associated with this observer.

The spacetime experienced by such a finite-lifetime observer defines the so-called \emph{causal diamond} (CD) spacetime. It can be geometrically characterized from the subset of Minkowski spacetime points restricted to the region $D_R$, denoted as $x^\mu = (t_{d_R}, x_{d_R})$. These coordinates are obtained by applying the conformal transformation introduced in Ref.~\cite{camblong2024entanglement} to the Minkowski coordinates restricted to the right Rindler wedge $R$ [see Fig.~\ref{rrw}], which are given by $\tilde{x}^\mu = (t_r, x_r)$. Accordingly, the conformal map is defined as
\begin{equation}
\tilde{x}^\mu \mapsto x^\mu = \qty[T \circ K \circ D]\, \tilde{x}^\mu.
\label{cm1}
\end{equation}

The transformation introduced in Eq.~\eqref{cm1} is given by the following finite conformal group transformations (see \cite{francesco2012conformal}):
\begin{align}
\text{Dilation:} \qquad & D(\lambda)\tilde{x}^\mu = \lambda\, \tilde{x}^\mu,\label{dktlbcdd}\\
\text{Translation:} \qquad & T(c)\tilde{x}^\mu = \tilde{x}^\mu + c^\mu,\label{dktlbctcc}\\
\text{Special Conformal Transformation (SCT):} \qquad & K(b)\tilde{x}^\mu = \frac{\tilde{x}^\mu - b^\mu (\tilde{x} \cdot \tilde{x})}{1 - 2(b \cdot \tilde{x}) + (b \cdot b)(\tilde{x} \cdot \tilde{x})}.
\label{dktlbc}
\end{align}
For later use, the parameters of the SCT and the translation are fixed as
\begin{equation}
b^\mu \equiv \left(0, -\frac{1}{2\alpha}\right)\quad, \quad 
c^\mu \equiv (0, -\alpha).
\label{pm1}
\end{equation}

Now, consider the coordinates $(t_r, x_r)$ defined through the following suitably rescaled transformation between Minkowski and Rindler coordinates $(\eta, \rho)$ (for a more detailed discussion, see Ref.~\cite{camblong2024entanglement}):
\begin{equation}
\frac{t_r}{\tilde{\alpha}} = \qty(\frac{2}{\alpha})\rho \sinh\left(\frac{2\eta}{\alpha}\right)\quad,\quad \frac{x_r}{\tilde{\alpha}} = \qty(\frac{2}{\alpha})\rho\cosh\left(\frac{2\eta}{\alpha}\right),\label{trrhoeta1}
\end{equation}
where $\tilde{\alpha} = 2\alpha/\lambda$. Then, by applying Eqs.~\eqref{cm1}–\eqref{pm1}, we obtain the coordinate transformation between the Minkowski points restricted to the region $D_R$ and the CD spacetime coordinates $(\eta, \rho)$
\begin{equation}
    t_{d_r} = \frac{4 \alpha^2 \rho \sinh \left(\frac{2 \eta }{\alpha} \right)}{\alpha^2 + 4\alpha\rho \cosh \left(\frac{2 \eta }{\alpha} \right) + 4\rho^2}\quad,\quad
    x_{d_r} = \frac{4\alpha\rho^2 - \alpha^3}{\alpha^2 + 4\alpha\rho \cosh \left(\frac{2 \eta }{\alpha} \right) + 4\rho^2},
\label{tdrxdretarho}
\end{equation}

It is worth noting that the dilation parameter $\lambda$ does not appear explicitly in Eq.~\eqref{tdrxdretarho}, as it has been absorbed into the rescaled coordinate transformation between Minkowski and Rindler spacetimes introduced in Eq.~\eqref{trrhoeta1}.\\

The line element of the CD spacetime takes the form
\begin{equation}
ds^2 = \Lambda^2(\eta,\rho) \left[ -\frac{4\rho^2}{\alpha^2} \, d\eta^2 + d\rho^2 \right], \label{cdstm1}
\end{equation}
where $\eta \in (-\infty, \infty)$ and $\rho > 0$, with the conformal factor given by
\begin{equation}
\Lambda^2(\eta,\rho) = \frac{16 \alpha^4}{\left( \alpha^2 + 4 \alpha \rho \cosh\left( \frac{2\eta}{\alpha} \right) + 4 \rho^2 \right)^2}.
\label{cdstm1cf}
\end{equation}

\subsection{Boundary of CD spacetime}\label{bbrdr}
\begin{figure}[h]
 \centering
   \begin{subfigure}{0.4\textwidth}
    \includegraphics[width=\linewidth]{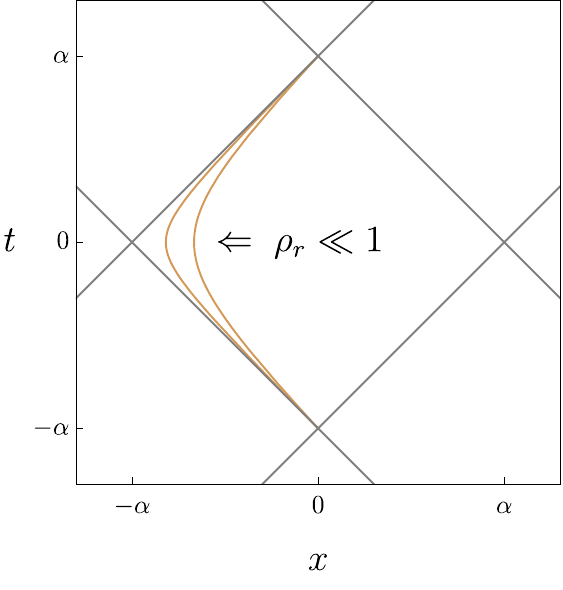}
     \caption{}\label{nhapfig}
   \end{subfigure}
   \hspace{0.5cm}
   \begin{subfigure}{0.4\textwidth}
     \centering
     \includegraphics[width=\linewidth]{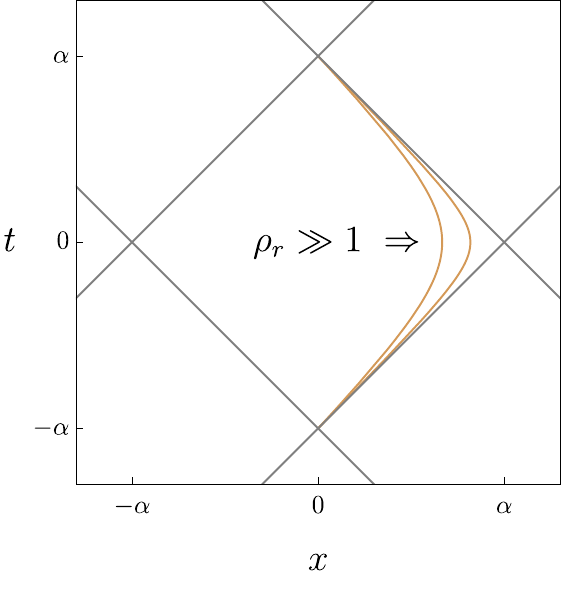}
     \caption{}\label{nhapfig2}
   \end{subfigure}
  \caption{Constant-$\rho$ trajectories in the causal diamond $D_R$ for (a) $\rho \ll 1$, and (b) $\rho \gg 1$.}
\end{figure}

Based on the coordinate transformation between Minkowski and the CD spacetime given in Eq.~\eqref{tdrxdretarho}, we observe that constant-$\rho$ trajectories in $D_R$ exhibit a characteristic behavior: for small values of $\rho$, these trajectories asymptotically approach the left boundary of the causal diamond $D_R$, while for large values of $\rho$, they approach the right boundary. These features are illustrated in Figs.~\ref{nhapfig} and \ref{nhapfig2}.\\

Such limiting behavior provides the basis for analyzing the boundary of the CD spacetime and its relation to the boundary of the region $D_R$. As we will show in this section, the boundary consists of the causal horizons, as well as the asymptotic regions in the CD spacetime, represented at the corners of $D_R$. This analysis will, in turn, enable us to describe the geometry in the vicinity of the causal horizons.

\subsubsection{Causal diamond horizons}
We begin by analyzing the expansion of $(t_{d_r}, x_{d_r})$ around the regime $\rho \ll 1$, which is given by
\begin{eqnarray}
t_{d_r} &=& 4 \rho \sinh\left(\frac{2 \eta}{\alpha}\right) - \frac{8 \rho^2 \sinh\left(\frac{4 \eta}{\alpha}\right)}{\alpha} + \frac{16 \rho^3 \sinh\left(\frac{6 \eta}{\alpha}\right)}{\alpha^2} + \mathcal{O}\qty[\qty(\rho/\alpha)^4],\label{r4rho1}\\
x_{d_r} &=& -\alpha + 4 \rho \cosh\left(\frac{2 \eta}{\alpha}\right) - \frac{8 \rho^2 \cosh\left(\frac{4 \eta}{\alpha}\right)}{\alpha} + \frac{16 \rho^3 \cosh\left(\frac{6 \eta}{\alpha}\right)}{\alpha^2} + \mathcal{O}\qty[\qty(\rho/\alpha)^4].\label{r4rho2}
\end{eqnarray}

\begin{figure}[h]
 \centering
   \begin{subfigure}{0.4\textwidth}
    \includegraphics[width=\linewidth]{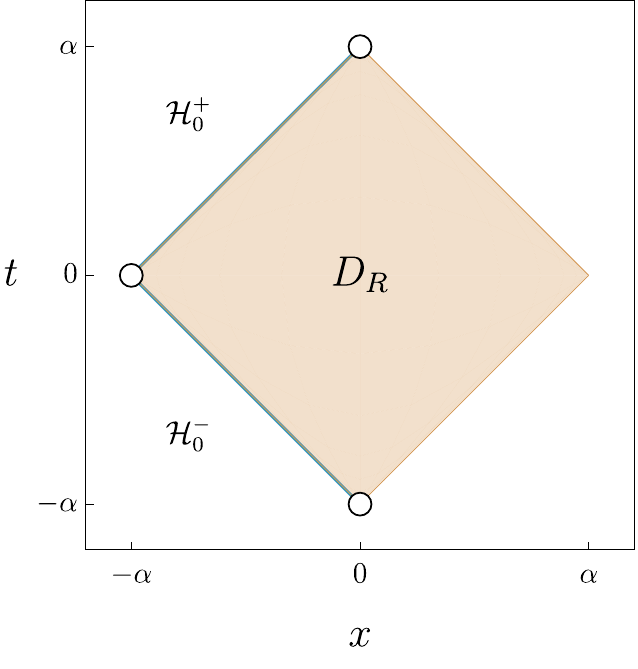}
     \caption{}\label{rho0infleft}
   \end{subfigure}
   \hspace{0.5cm}
   \begin{subfigure}{0.4\textwidth}
     \centering
     \includegraphics[width=\linewidth]{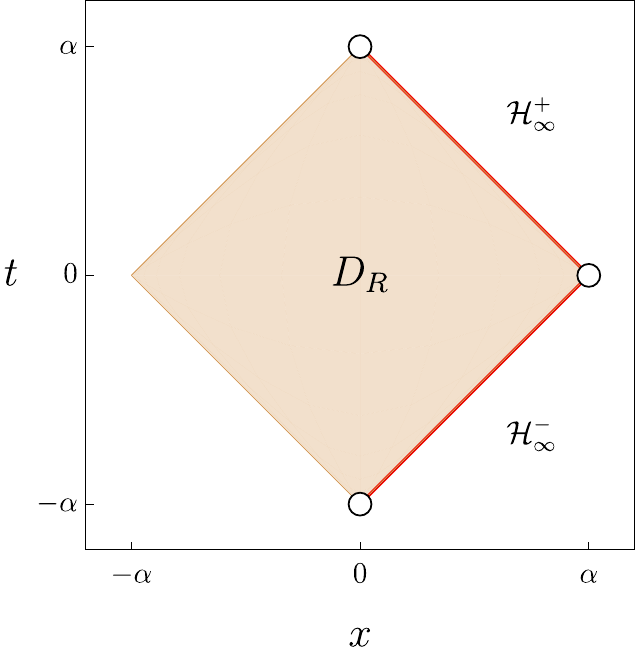}
     \caption{}\label{rho0infright}
   \end{subfigure}
  \caption{(a) causal horizons at the left boundary of $D_R$. (b) causal horizons at the right boundary of $D_R$.}
\end{figure}

In addition, since $\sinh(2\eta/\alpha) \approx \pm \cosh(2\eta/\alpha)$ for $\eta \gg 1$ and $-\eta \gg 1$, respectively, it follows from Eqs.~\eqref{r4rho1} and \eqref{r4rho2} that, in the simultaneous limits $\rho \to 0$ and $\eta \to \pm\infty$, one directly obtains the following causal horizons [see Fig.~\ref{rho0infleft}]
\begin{eqnarray}
\mathcal{H}^+_0&:& \quad t_{d_r} = x_{d_r} + \alpha \quad,\quad \rho\to0\;,\;\eta \to \infty,\\
\mathcal{H}^-_0&:& \quad t_{d_r} = -x_{d_r} - \alpha \quad,\quad \rho\to0\;,\;\eta \to -\infty,
\end{eqnarray}
where $-\alpha < x < 0$. This range reflects the fact that both limits are taken simultaneously, while the precise values at $x=0$ and $x=-\alpha$ will be analyzed in detail in the study of the corners of $D_R$.\\

On the other hand, we now turn to the expansion of $(t_{d_r}, x_{d_r})$ in the regime $\rho \gg 1$, from which we obtain
\begin{eqnarray}
t_{d_r} &=& \frac{\alpha^2 \sinh\left(\frac{2 \eta}{\alpha}\right)}{\rho} - \frac{\alpha^3 \sinh\left(\frac{4 \eta}{\alpha}\right)}{2 \rho^2} + \frac{\alpha^4 \sinh\left(\frac{6 \eta}{\alpha}\right)}{4 \rho^3} + \mathcal{O}\qty[\qty(\rho/\alpha)^{-4}],\label{rhoinf1}\\
x_{d_r} &=& \alpha - \frac{\alpha^2 \cosh\left(\frac{2 \eta}{\alpha}\right)}{\rho} + \frac{\alpha^3 \cosh\left(\frac{4 \eta}{\alpha}\right)}{2 \rho^2} - \frac{\alpha^4 \cosh\left(\frac{6 \eta}{\alpha}\right)}{4 \rho^3} + \mathcal{O}\qty[\qty(\rho/\alpha)^{-4}].\label{rhoinf2}
\end{eqnarray}

As in the previous case, applying the limits $\rho \to \infty$ and $\eta \to \pm\infty$ to Eqs.~\eqref{rhoinf1} and \eqref{rhoinf2} yields the following causal horizons [see Fig.~\ref{rho0infright}]
\begin{eqnarray}
\mathcal{H}^+_\infty &:& \quad t_{d_r} = -x_{d_r} + \alpha \quad,\quad \rho\to\infty\;,\;\eta \to \infty,\\
\mathcal{H}^-_\infty &:& \quad t_{d_r} = x_{d_r} - \alpha \quad,\quad \rho\to\infty\;,\;\eta \to -\infty,
\end{eqnarray}
where $0 < x_{d_r} < \alpha$.

\subsubsection{Causal diamond corners}
The asymptotic regions of the CD spacetime corresponding to the four corners of $D_R$ can be identified by analyzing the coordinate transformations between Minkowski and the CD spacetime in the appropriate limits
\begin{eqnarray}
i^0&:& \quad(t_{d_r},x_{d_r})=(0,-\alpha)\quad,\quad\rho\to 0,\eta\in(-\infty,\infty),\\
i^\infty&:&\quad (t_{d_r},x_{d_r})=(0,\alpha)\quad,\quad\rho\to \infty,\eta\in(-\infty,\infty),\\
i^-&:&\quad (t_{d_r},x_{d_r})=(-\alpha,0)\quad,\quad\rho\in(0,\infty),\eta\to-\infty,\\
i^+&:&\quad (t_{d_r},x_{d_r})=(\alpha,0)\quad,\quad\rho\in(0,\infty),\eta\to\infty.
\end{eqnarray}

\begin{figure}[h]
 \centering
   \begin{subfigure}{0.35\textwidth}
    \includegraphics[width=\linewidth]{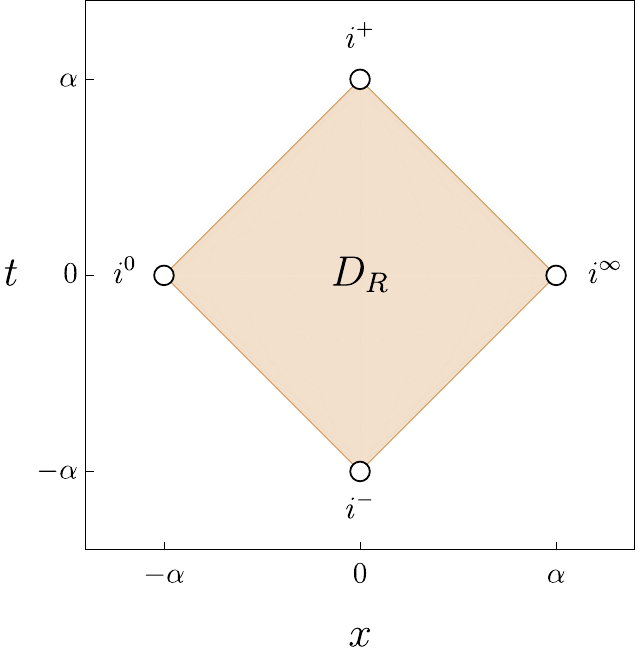}
     \caption{}\label{figp4}
   \end{subfigure}
   \hspace{0.5cm}
   \begin{subfigure}{0.45\textwidth}
     \centering
     \includegraphics[width=\linewidth]{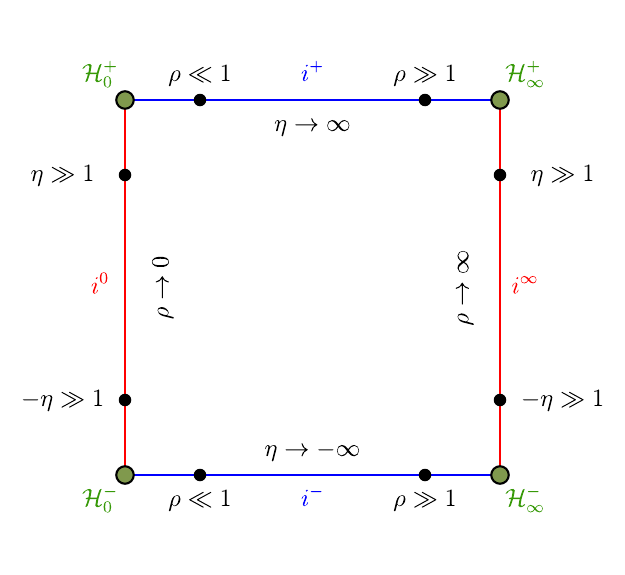}
     \caption{}\label{fdigc}
   \end{subfigure}
  \caption{(a) asymptotic regions of the CD spacetime represented by the four corners of the causal diamond $D_R$. (b) representation of the extended view of the corners in $D_R$}
\end{figure}

In Fig.~\ref{figp4}, we show the labels corresponding to the corners of $D_R$, which correspond to the analyzed asymptotic regions. Likewise, Fig.~\ref{fdigc} shows an “extended” representation, providing a clearer view of these regions.

\subsection{Effective surface gravity of the CD spacetime horizons}\label{surfg}

\begin{figure}[h]
 \centering
   \begin{subfigure}{0.4\textwidth}
    \includegraphics[width=\linewidth]{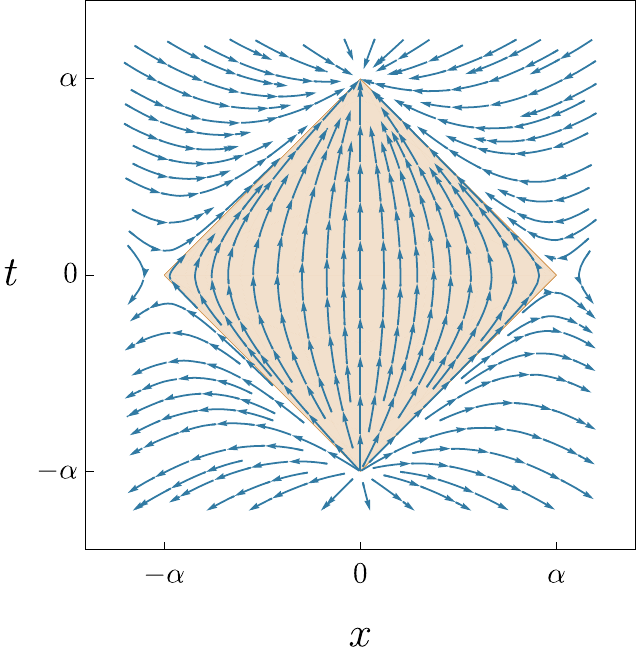}
     \caption{}\label{kvcd}
   \end{subfigure}
   \hspace{0.5cm}
   \begin{subfigure}{0.4\textwidth}
     \centering
     \includegraphics[width=\linewidth]{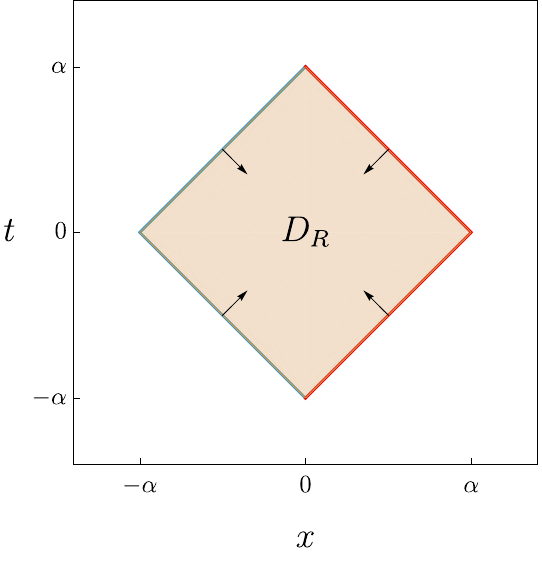}
     \caption{}\label{nvcd}
   \end{subfigure}
   \caption{(a) Geometric flow generated by the conformal Killing vector $\partial_\eta$ in the extended Minkowski geometry. (b) Normal vectors to the horizons of the region $D_R$.}
\end{figure}

In the CD spacetime geometry, the metric depends explicitly on the temporal coordinate $\eta$. As a result, the differential operator $\chi^{(\eta)} = \partial_\eta$ is not a Killing vector field, but rather a conformal Killing vector, whose geometric flow is illustrated in Fig.~\ref{kvcd}. In Minkowski coordinates, this vector takes the form
\begin{equation}
\chi^{(\eta)} = \left(\frac{\alpha^2 - t^2 - x^2}{\alpha^2}\right) \partial_t - \left(\frac{2 t x}{\alpha^2} \right) \partial_x.
\label{confamin}
\end{equation}

Likewise, noting that the boundary of the CD spacetime corresponds, in Minkowski spacetime, to the boundary of the region $D_R$, given by $\partial D_R:=\qty{(t,x): \lvert t \rvert + \lvert x \rvert = \alpha}$, we observe that it precisely corresponds to the Killing horizon of the vector field $\chi^{(\eta)}$
\begin{equation}
    \eval{\chi^{(\eta)} \cdot \chi^{(\eta)}}_{\partial D_R} = 0.
    \label{ckvh1234}
\end{equation}

Since $\chi^{(\eta)} \cdot \chi^{(\eta)} = 0$ on the boundary of $D_R$, the gradient of this norm must be proportional to the normal vector to the Killing horizon, which in this case coincides with the conformal Killing vector itself. As shown in Ref.~\cite{jacobson1993conformal}, the proportionality factor defines an effective surface gravity $\kappa$, which can be associated with a Hawking-like temperature given by $T_H = \kappa/(2\pi)$. It is important to emphasize that, in the present context, we are not dealing with gravitational fields or black hole geometry. Therefore, this surface gravity should be understood as an effective parameter generating an apparent Hawking temperature.\\

Thus, the effective surface gravity associated with the conformal Killing vector $\chi^{(\eta)}$ is defined by evaluating the following identity on the boundary of $D_R$
\begin{equation}
\nabla^\mu \left[\chi^{(\eta)} \cdot \chi^{(\eta)}\right] = 2 \hat{\epsilon} \, \kappa \, \chi^\mu_{(\eta)},
\label{sugrckvv1}
\end{equation}
where $\hat{\epsilon} = \pm 1$ depending on whether the normal vector to a given segment of the boundary of $D_R$ is future-directed $(+)$ or past-directed $(-)$ [see Fig.~\ref{nvcd}].\\

Finally, by evaluating Eq.~\eqref{sugrckvv1} explicitly in Minkowski coordinates—using the expression for $\chi^{(\eta)}$ given in Eq.~\eqref{confamin} together with the relation $\chi^{(\eta)} \cdot \chi^{(\eta)} = \eta_{ab} \chi^a_{(\eta)} \chi^b_{(\eta)}$—we obtain the following value for the effective surface gravity and its corresponding Hawking-like temperature
\begin{equation}
    \kappa = \frac{2}{\alpha} \quad \Rightarrow \quad T_H = \frac{\kappa}{2\pi} = \frac{1}{\pi\alpha} = T_D.
    \label{sufgrv1}
\end{equation}

Thus, the obtained temperature exactly matches the temperature $T_D$ associated with the thermal description of the Minkowski vacuum as perceived by an observer with finite lifetime $2\alpha$, restricted to the causal diamond $D_R$. This result is consistent with previous findings based on the thermal time hypothesis, the open quantum system approach, the thermofield double construction, and related frameworks~\cite{martinetti2003diamond,chakraborty2022thermal,camblong2024conformal,chakraborty2024path,su2016spacetime,foo2020generating,foo2025superpositions}. In what follows, we adopt this identification of $T_D$ in terms of the surface gravity $\kappa$, namely $T_D = \kappa / (2\pi) = (\pi \alpha)^{-1}$.

\subsection{Geodesics in the spacetime of the causal diamond}
\subsubsection{Geodesic trajectories}
From the preceding geometric analysis, we conclude that, when viewed from Minkowski spacetime, the causal diamond $D_R$ exhibits a structure analogous to that of a Penrose-Carter diagram. This correspondence follows from the fact that the CD spacetime metric, given in Eq.~\eqref{cdstm1}, is conformally related to the Minkowski metric through the coordinate transformation $\rho = e^{2\xi/\alpha}$.\\

\begin{figure}[h]
 \centering
  \begin{subfigure}{0.26\textwidth}
    \includegraphics[width=\linewidth]{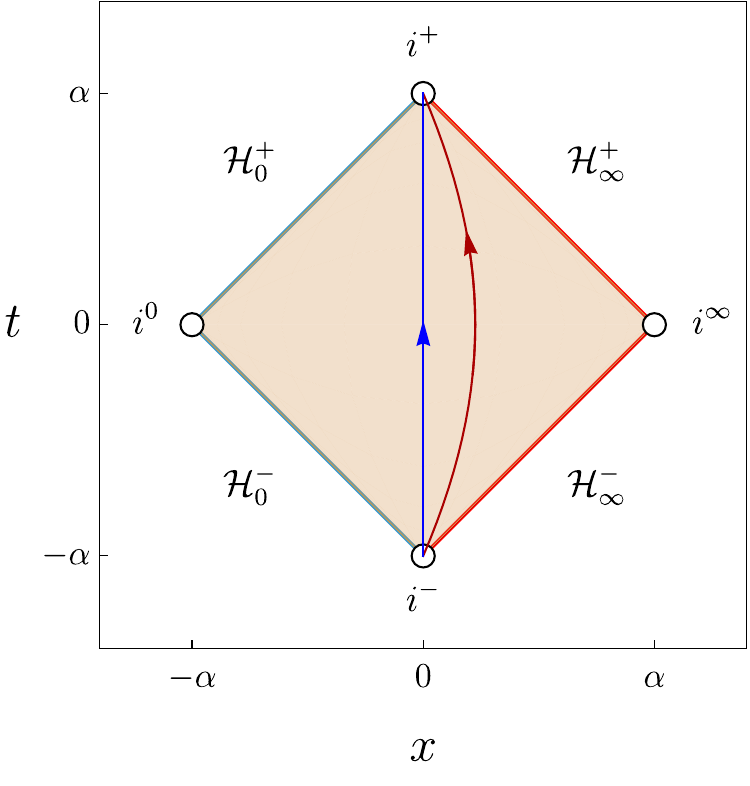}
    \caption{}\label{geocd1}
  \end{subfigure}
  \hspace{0.5cm}
  \begin{subfigure}{0.35\textwidth}
    \includegraphics[width=\linewidth]{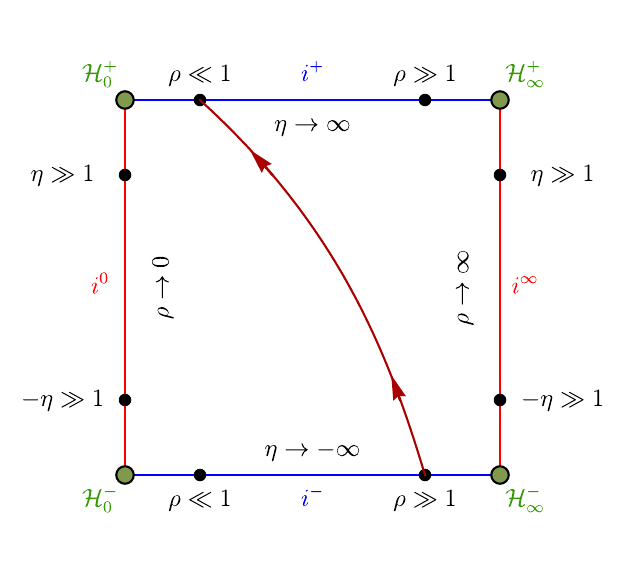}
    \caption{}\label{geoex}
  \end{subfigure}
  \hspace{0.5cm}
  \begin{subfigure}{0.26\textwidth}
    \includegraphics[width=\linewidth]{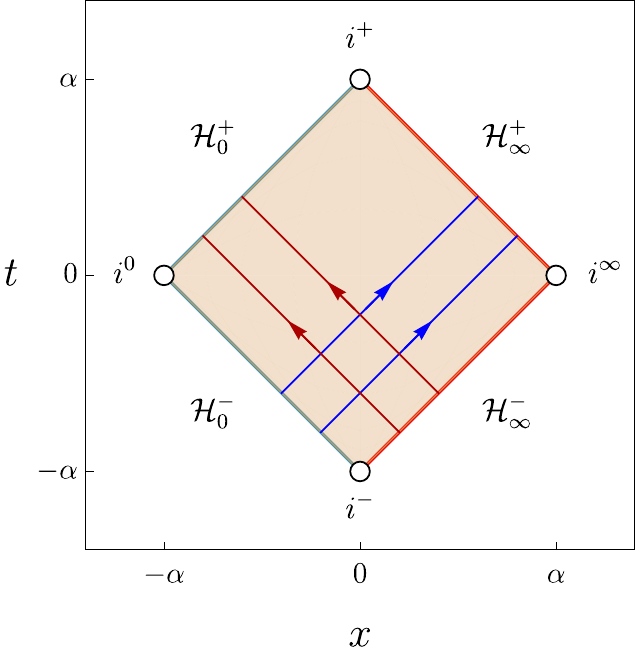}
    \caption{}\label{geocd2}
  \end{subfigure}
  \caption{(a) timelike geodesics. (b) a timelike geodesic in the extended representation of $i^-$ and $i^+$. (c) Null geodesics.}\label{fig666beast}
\end{figure}

Given this conformal relation, the causal structure is preserved in both spacetimes. Accordingly, the causal horizons $\mathcal{H}^-_0$ and $\mathcal{H}^-_\infty$ are null surfaces from which null trajectories can only emerge, whereas $\mathcal{H}^+_0$ and $\mathcal{H}^+_\infty$ are null surfaces where null trajectories can only terminate, while $i^-$ and $i^+$ represent past and future timelike infinities, respectively, where timelike geodesics begin and end. These geodesic trajectories are depicted in Fig.~\ref{fig666beast}.\\

\subsubsection{Geodesics in the near-horizon approximation}\label{geonhap1}
As can be seen from Eqs.~\eqref{r4rho1} and \eqref{r4rho2}, in the leading-order approximation for $\rho \ll 1$, the CD spacetime geometry is characterized by an effective metric that is independent of $\eta$
\begin{equation}
    ds^2_{\rho\ll 1} \sim 16\left(-\frac{4}{\alpha^2} \rho^2 d\eta^2 + d\rho^2\right).
    \label{mchr01}
\end{equation}

To investigate the behavior of timelike geodesics in the near-horizon approximation $\rho \ll 1$, we must determine the functional dependence of $\eta$ on $\rho$, as well as the proper time $\tau$. To this end, we consider a freely falling massive particle in the effective geometry described by the metric $ds^2_{\rho\ll 1}$. Owing to the static nature of this metric, there exists a conserved quantity along geodesics—an effective energy per unit mass—given by
\begin{equation}
\tilde{e} = -g_{\eta\eta} \frac{d\eta}{d\tau} \quad \Rightarrow \quad \frac{d\eta}{d\tau} = \frac{e \alpha^2}{64\rho^2}, 
\label{geqffo122}
\end{equation}
where $d\tau$ denotes the proper time experienced by the freely falling particle.\\

Then, using the expression for the effective energy in Eq.~\eqref{geqffo122} and noting that proper time is defined via $ds^2_{\rho\ll 1} = -d\tau^2$, we obtain
\begin{equation}
\frac{d\rho}{d\tau} = \mp\frac{1}{4}\left(\frac{\tilde{e}^2 \alpha^2}{64 \rho^2} - 1\right)^{1/2}, \quad
\frac{d\eta}{d\rho} = \mp\frac{\tilde{e} \alpha^2}{16 \rho^2} \left(\frac{\tilde{e}^2 \alpha^2}{64 \rho^2} - 1 \right)^{-1/2}, \label{drhodlmb1}
\end{equation}
where the $(-)$ sign corresponds to motion toward the future, for which $\rho$ decreases, while the $(+)$ sign corresponds to motion toward the past, also implying a decreasing $\rho$ in this context.\\

Finally, considering once again the approximation $\rho \ll 1$, we find the asymptotic behavior of $\eta$ and $\tau$ as functions of $\rho$
\begin{eqnarray}
\eta = \mp\frac{1}{\kappa} \ln \rho + \mathcal{O}(\rho^2) + \text{const.}, \quad
\tau = \mp\frac{8 \kappa}{\tilde{e}} \rho^2 + \mathcal{O}(\rho^4) + \text{const.},\label{asympcb}
\end{eqnarray}
where we have identified the effective surface gravity constant $\kappa = 2/\alpha$, as obtained in Eq.~\eqref{sufgrv1}. Finally, Eq.~\eqref{asympcb} shows that for $\rho \ll 1$, one finds $\eta \to \pm \infty$, which corresponds precisely to the regions near the horizons $\mathcal{H}^+_0$ and $\mathcal{H}^-_0$, contained in the $i^+$ and $i^-$ regions, respectively, specifically very close to the left boundary, as shown in Fig.~\ref{fdigc}.\\

Analogously to the previous case, the CD spacetime geometry at leading order approximation for $\rho\gg 1$ exhibits an effective metric that is independent of $\eta$ [see Eqs. \eqref{rhoinf1} and \eqref{rhoinf2}]
\begin{equation}
ds^2_{\rho \gg 1} \sim \frac{\alpha^4}{\rho^4}\left(-\frac{4}{\alpha^2} \rho^2 d\eta^2 + d\rho^2\right).\label{mchr012}
\end{equation}

Interestingly, by defining $4\tilde{\rho} = \alpha^2 / \rho$, which corresponds to the regime $\tilde{\rho} \ll 1$, the metric reduces to the same form as that obtained in the case $\rho \ll 1$
\begin{equation}
ds^2_{\rho \gg 1} = ds^2_{\tilde{\rho} \ll 1}\sim 16\left(-\frac{4}{\alpha^2} \tilde{\rho}^2 d\eta^2 + d\tilde{\rho}^2\right).\label{mchr0123}
\end{equation}

Proceeding analogously to the case of $\rho \ll 1$, we find that in the regime $\tilde{\rho} \ll 1$, the asymptotic behavior of $\eta$ and $\tau$ as a function of $\tilde{\rho}$ is given by
\begin{eqnarray}
\eta = \mp\frac{1}{\kappa} \ln \tilde{\rho} + \mathcal{O}(\tilde{\rho}^2) + \text{const.} \quad,\quad 
\tau = \mp\frac{8 \kappa}{\tilde{e}} \tilde{\rho}^2 + \mathcal{O}(\tilde{\rho}^4) + \text{const.},\label{etanadkambdar1}
\end{eqnarray}
where, for $\tilde{\rho} \ll 1$, one has $\eta \to \pm \infty$, which corresponds precisely to the regions near the horizons $\mathcal{H}^+_\infty$ and $\mathcal{H}^-_\infty$, contained in the $i^+$ and $i^-$ regions, respectively, specifically very close to the right boundary, as shown in Fig.~\ref{fdigc}.

\section{Field Modes in the Near-Horizon Approximation and Conformal Quantum Mechanics}\label{seciii}
The equation of motion for a real massive scalar field in a spacetime with metric tensor $g_{\mu\nu}$ is given by
\begin{equation}
    \left[\frac{1}{\sqrt{-g}}\, \partial_\mu \left(\sqrt{-g}\, g^{\mu\nu} \partial_\nu\right) - m^2 - \xi R\right] \phi = 0.
\end{equation}

In the case of the causal diamond geometry, the metric is described by the line element and conformal factor given in Eqs.~\eqref{cdstm1} and \eqref{cdstm1cf}, respectively. Since the Ricci scalar vanishes in this spacetime ($R = 0$), the equation of motion simplifies to
\begin{equation}
\left[-\partial^2_\eta + \frac{4\rho}{\alpha^2} \partial_\rho + \frac{4\rho^2}{\alpha^2} \partial^2_\rho - \frac{4\rho^2}{\alpha^2}\Lambda^2(\eta,\rho)m^2 \right] \phi(\eta,\rho) = 0.
\label{eom123}
\end{equation}

As seen from Eq.~\eqref{eom123}, the conformal factor $\Lambda^2(\eta,\rho)$, defined in Eq.~\eqref{cdstm1cf}, prevents a straightforward separation of variables due to its nontrivial dependence on both $\eta$ and $\rho$. Nevertheless, insight into the behavior of the field modes can be gained by analyzing the asymptotic properties of the conformal factor near the boundaries of the CD spacetime.\\

As discussed in Sec.~\ref{bbrdr}, around the regimes $\rho \ll 1$ and $\rho \gg 1$, the conformal factor $\Lambda^2$ exhibits the following asymptotic behavior
\begin{eqnarray}
\rho \ll 1: &&\quad \Lambda^2(\eta,\rho) \sim 16 + \mathcal{O}\qty[(\rho/\alpha)^2], \\
\rho \gg 1: &&\quad \Lambda^2(\eta,\rho) \sim \frac{\alpha^4}{\rho^4} + \mathcal{O}\qty[(\rho/\alpha)^{-6}],
\end{eqnarray}
which precisely correspond to the effective near-boundary metrics given in Eqs.~\eqref{mchr01} and \eqref{mchr012}.\\

Then, in the approximation $\rho \ll 1$, the term $(4\rho^2/\alpha^2)\Lambda^2 m^2$ in Eq.~\eqref{eom123} reduces to $(4\rho^2/\alpha^2)16m^2$. Thus, near the causal horizons $\mathcal{H}^+_0$ and $\mathcal{H}^-_0$, the field equation becomes
\begin{eqnarray}
\left[-\partial^2_\eta + \frac{4\rho}{\alpha^2} \partial_\rho + \frac{4\rho^2}{\alpha^2} \partial^2_\rho - \frac{4\rho^2}{\alpha^2}(16m^2) \right] \phi(\eta,\rho) = 0. \label{eomnh1}
\end{eqnarray}

This motivates the following ansatz for the field modes:
\begin{equation}
    \phi_\omega(\eta,\rho) \sim e^{-i\omega\eta} \frac{\psi(\rho)}{\sqrt{\rho}}, \label{anzets}
\end{equation}
where the prefactor $1/\sqrt{\rho}$ facilitates a Liouville-type transformation of the radial equation. Substituting this ansatz into Eq.~\eqref{eomnh1}, we obtain the following effective radial equation:
\begin{equation}
\psi^{\prime\prime}(\rho) + \frac{1}{\rho^2} \qty(\frac{1}{4} + \Theta^2)\psi(\rho) \approx 0, \label{cqmeqrho0}
\end{equation}
where we have defined the conformal parameter $\Theta = \omega\alpha/2$, which can be expressed in terms of the effective surface gravity $\kappa = 2/\alpha$ obtained in Eq.~\eqref{sufgrv1}. In this way, the conformal factor can be written as $\Theta = \omega/\kappa$. Furthermore, we note that, in this near-horizon approximation, the mass term becomes negligible and therefore does not appear in the reduced equation.\\

Similarly, in the asymptotic regime $\rho \gg 1$, which corresponds to the region near the causal horizons $\mathcal{H}^+_\infty$ and $\mathcal{H}^-_\infty$, the conformal factor scales as $\Lambda^2 \sim \alpha^4/\rho^4$. As a result, the term $4\rho^2/(\alpha^2 \Lambda^2)$ scales as $\mathcal{O}[(\rho/\alpha)^{-2}]$ and becomes subdominant. Adopting the same ansatz as in Eq.~\eqref{anzets}, we again find that the mass term is suppressed, and the field equation reduces to the same expression as in Eq.~\eqref{cqmeqrho0}.\\

Thus, in both asymptotic regimes, $\rho \ll 1$ and $\rho \gg 1$, the field equation reduces to a Schrödinger-like eigenvalue problem characteristic of conformal quantum mechanics (CQM), a one-dimensional theory with $\mathfrak{sl}(2,\mathbb{R})$ symmetry (see Ref. \cite{de1976conformal,camblong2023spectral,arzano2020conformal}).\\

The solution to the near-horizon ($\rho \ll 1$) CQM equation, given by Eq.~\eqref{cqmeqrho0}, is  
\begin{equation}
\frac{\psi(\rho)}{\sqrt{\rho}} \sim \rho^{\pm i\Theta},\label{cqmeom11}
\end{equation}
where $(+)$ corresponds to outgoing modes (away from the horizon) and $(-)$ to ingoing modes (toward the horizon). Consequently, the near-horizon field modes with positive frequency take the form of $(\pm)$ CQM modes
\begin{equation}
    \phi_\omega^{\pm\,(\mathrm{CQM})}(\eta,\rho) \sim e^{-i\omega\eta}\rho^{\pm i\Theta}.\label{cqmmodf1}
\end{equation}
It is also important to note that in the near-horizon region with $\rho \gg 1$, the solutions of the CQM equation take the same form, but now the $(+)$ sign corresponds to ingoing modes, whereas the $(-)$ sign corresponds to outgoing modes.

\section{HBAR in Causal Diamond Geometry}
\subsection{Setup: Atom–field interaction}
The physical system under consideration consists of a quantum scalar field and a cloud of two-level atoms, interacting weakly along the atoms' trajectories. This setup was originally proposed in Ref.~\cite{scully2018quantum}, which introduced the phenomenon of HBAR. Here, we apply this framework to the CD spacetime geometry, which distinguishes our study from previous works focusing on black hole geometries~\cite{camblong2020near,azizi2021acceleration,azizi1,azizi2}.\\

We begin by introducing a real scalar field operator $\hat{\phi}(\eta, \rho)$ defined on the causal diamond spacetime [characterized by the metric in Eq.~\eqref{cdstm1}], together with a well-defined Boulware vacuum state $\ket{0_B}$—understood here as a generalization of the Boulware vacuum originally defined in the context of black hole geometries in Ref.~\cite{boulware1975quantum}. This quantization scheme is justified by the fact that the CD spacetime metric is conformally related to Minkowski spacetime, and therefore admits a global Cauchy surface, ensuring global hyperbolicity. As a result, the scalar field can be consistently quantized with suitable boundary conditions at the causal boundaries and expanded in terms of a complete orthonormal set of mode solutions (see Ref. \cite{camblong2024entanglement}). In the free theory, the field expansion takes the form:
\begin{equation}
    \hat{\phi}(\eta, \rho) = \sum_{\mathbf{s}}\left[\hat{a}_{\mathbf{s}}\, \phi_{\mathbf{s}}(\eta,\rho) + \hat{a}^{\dagger}_{\mathbf{s}}\, \phi^*_{\mathbf{s}}(\eta,\rho)\right],
\end{equation}
where $\hat{a}_{\mathbf{s}}$ is the annihilation operator associated with mode $\phi_{\mathbf{s}}$ and satisfying $\hat{a}_{\mathbf{s}} \ket{0_B} = 0$, and $\mathbf{s}$ labels the quantum numbers characterizing each mode.\\

The other component of the system consists of a cloud of atoms, modeled as two-level systems. In this work, we focus on the process in which atoms are injected at random times from the asymptotic past, $\eta \to -\infty$, at $\rho \gg 1$, and subsequently follow geodesic trajectories toward the causal horizon $\mathcal{H}^+_0$ (the analogous injection process from $\eta \to -\infty$ at $\rho \ll 1$, directed toward $\mathcal{H}^+_\infty$, is discussed in Appendix~\ref{ap12}). Being massive, the atoms eventually reach the asymptotic future, $\eta \to \infty$, terminating at $\rho \ll 1$ within the region $i^+$, close to the left corner, as illustrated in Fig.~\ref{geoex}. Furthermore, while moving along these free-fall trajectories, the atoms interact with the quantum field. Although the field remains in its vacuum state, the freely falling atoms perceive it differently due to their relative acceleration with respect to the field modes. As a consequence, the
exchange of virtual particles is disrupted, leading to spontaneous emission and absorption processes in which the atoms interact with scalar field quanta (see Refs. \cite{scully2018quantum,azizi1}).\\

The interaction between each atom and the field is modeled by the dipole interaction Hamiltonian in the interaction picture
\begin{equation}
V_I(\tau) = g \left[\hat{a}_{\mathbf{s}}\, \phi_{\mathbf{s}}(\eta(\tau),\rho(\tau)) + \text{H.c.} \right] \left( \hat{\sigma}\, e^{-i\nu \tau} + \text{H.c.} \right),\label{intvi}
\end{equation}
where H.c. stands for the Hermitian conjugate operation, $g$ is the coupling constant, and $\hat{\sigma}$ is the atomic lowering operator corresponding to a transition of frequency $\nu$. The interaction is evaluated along the atom's worldline, parametrized by proper time $\tau$.\\

Considering the initial state of the atom–field system given by $\ket{0_B, b}$, with $\ket{0_B}$ the Boulware vacuum and $\ket{b}$ the atomic ground state, the interaction in Eq.~\eqref{intvi} allows the atom, while in free fall, to become excited and emit a scalar quantum via the operator $\hat{a}^\dagger_{\mathbf{s}} \hat{\sigma}^\dagger$ (see Ref. \cite{PhysRevLett.91.243004}). For sufficiently small $g$, we can apply first-order perturbation theory, from which we obtain that the transition probability to the final state $\ket{1_{\mathbf{s}}, a}$—where $\ket{1_{\mathbf{s}}}$ is a one-particle field state and $\ket{a}$ is the excited atomic state—is
\begin{equation}
    P_{\rm em,\mathbf{s}} = \left|\int d\tau\, \bra{1_{\bf s}, a} V_I(\tau) \ket{0_B, b}\right|^2 = g^2 \left|\int d\tau\, \phi^*_{\textbf{s}}(\eta(\tau), \rho(\tau))\, e^{i\nu\tau} \right|^2.\label{probem1}
\end{equation}

Similarly, another free-falling atom may absorb the emitted scalar quantum, causing the system to transition from $\ket{1_{\mathbf{s}}, b}$ to $\ket{0_B, a}$ via the operator $\hat{a}_{\mathbf{s}} \hat{\sigma}^\dagger$. The corresponding absorption probability is
\begin{equation}
    P_{\rm ab,\mathbf{s}} = \left|\int d\tau\, \bra{0_B, a} V_I(\tau) \ket{1_{\bf s}, b} \right|^2 = g^2 \left|\int d\tau\, \phi_{\textbf{s}}(\eta(\tau), \rho(\tau))\, e^{i\nu\tau} \right|^2.
\end{equation}
\subsection{Emission and absorption probabilities}\label{emabprb1}
\begin{figure}[h]
\centering
\includegraphics[width=0.5\linewidth]{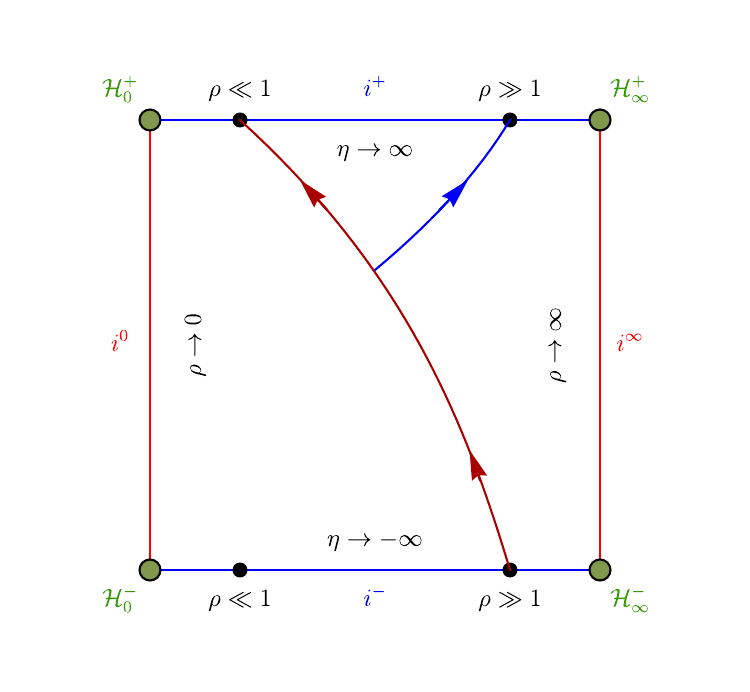}
\caption{Representation of a freely falling atom approaching $\mathcal{H}^+_0$ (dark red curve) and the emitted scalar quantum (blue curve).}\label{frefall1}
\end{figure}

The emission and absorption probabilities of scalar quanta by freely falling atoms near the causal horizon $\mathcal{H}^+_0$ can be obtained from the near-horizon analysis. This requires the near-horizon behavior of $\eta$ and $\tau$, as given in Eq.~\eqref{asympcb}, together with the CQM modes in Eq.~\eqref{cqmmodf1}, where the outgoing $(+)$ mode corresponds to quanta moving away from the horizon toward the region $\rho \gg 1$, as represented in Fig. \ref{frefall1} while the ingoing $(-)$ mode corresponds to quanta moving toward the horizon.\\

Given our interest in characterizing the physical nature of these processes, it is more convenient to compute the corresponding rates of emission and absorption, defined as $R_{\rm em,\mathbf{s}} = \mathfrak{r} P_{\rm em,\mathbf{s}}$ and $R_{\rm ab,\mathbf{s}} = \mathfrak{r} P_{\rm ab,\mathbf{s}}$, respectively. Here, $\mathfrak{r} = \Delta N / \Delta \eta$ denotes the atom injection rate, with $\Delta N$ representing the number of atoms injected during the interval $\Delta \eta$ (see Ref.~\cite{azizi1} for further details). In the following, we focus on evaluating the emission rate near the causal horizon $\mathcal{H}^+_0$ (i.e., for $\rho \ll 1$), analyzing the integral characterizing this process as given in Eq.~\eqref{probem1} for the outgoing $(+)$ CQM mode
\begin{eqnarray}
\int d\tau\, [\phi^{+(\text{CQM})}_{\omega}(\eta(\tau), \rho(\tau))]^* e^{i\nu\tau} 
&=& \int d\tau\, e^{i\omega\eta} \rho^{-i\Theta} e^{i\nu\tau} \nonumber\\
&=& \frac{16\kappa}{\tilde{e}} \int_0^{\rho_\circ} d\rho\, \rho\,\rho^{-2i\omega/\kappa} e^{- i8\nu\kappa\rho^2/\tilde{e}},
\label{eqintm1}
\end{eqnarray}
where the upper limit of the integral, $\rho_\circ \ll 1$, implements the constraint on $\rho$.\\

The behavior of this integral is governed by the interplay between the two oscillatory factors, $\rho\,\rho^{-2i\omega/\kappa}$ and $e^{-i8\nu\kappa\rho^2/\tilde{e}}$. The first factor can be decomposed into a logarithmically oscillating term, $\rho^{-2i\omega/\kappa}$—which originates from the conformal quantum mechanics (CQM) structure of the near-horizon region—and a linear prefactor $\rho$, which suppresses the amplitude while remaining finite for $\rho > 0$. The scale invariance of $\rho^{-2i\omega/\kappa}$, characterized by the dimensionless conformal parameter $\Theta = \omega/\kappa$, makes it the dominant contribution in the regime $\rho \ll 1$, where the exponential factor $e^{-i8\nu\kappa\rho^2/\tilde{e}}$ is approximately unity. As $\rho$ increases away from the horizon, the exponential term induces rapid oscillations. Thus, these oscillations average out and contribute negligibly to the integral. Consequently, the upper limit $\rho_\circ$ in Eq.~\eqref{eqintm1} can be effectively extended to infinity
\begin{align}
\int d\tau\, \phi^*_\omega(\eta(\tau),\rho(\tau))\, e^{i\nu\tau}
&= \frac{16\kappa}{\tilde{e}} \int_0^{\infty} d\rho\, \rho \rho^{-2i\omega/\kappa} e^{-i8\nu\kappa\rho^2/e} \nonumber\\
&=\frac{-i 2^{3 i \omega /\kappa } }{\nu }\Gamma \left(1-\frac{i \omega }{\kappa }\right) \left(\frac{i \kappa  \nu }{\tilde{e}}\right)^{i \omega/\kappa },
\end{align}
where, in order to regularize the integral, we have introduced a small positive parameter $\epsilon > 0$ through the substitution $\nu \to \nu - i\epsilon$, and finally taken the limit $\epsilon \to 0^+$.\\

The emission rate then becomes
\begin{equation}
R_{\text{em},\mathbf{s}} = \frac{2\pi \mathfrak{r} g^2 \omega}{\kappa \nu^2} \left( \frac{1}{e^{2\pi\omega/\kappa} - 1} \right).
\label{emrate}
\end{equation}

The absorption rate $R_{\text{ab}}$ can be obtained analogously, or directly from the emission rate by applying the substitution $\omega \to -\omega$, yielding
\begin{equation}
R_{\text{ab},\mathbf{s}} = \frac{2\pi \mathfrak{r} g^2 \omega}{\kappa \nu^2} \left( \frac{1}{1 - e^{-2\pi \omega/\kappa}} \right) = e^{2\pi\omega/\kappa} R_{\text{em},\mathbf{s}}.
\label{remrabs12}
\end{equation}

At this point it is important to note that calculations of the emission and absorption rates using the ingoing $(-)$ CQM modes lead to the cancellation of the logarithmic phases, which ultimately results in both processes having zero probability (as shown in Ref.~\cite{camblong2020near} for the case of a black hole geometry). Therefore, henceforth the emission and absorption rates should be understood as always referring to the outgoing $(+)$ CQM modes.

\subsection{Thermal behavior: Radiation field density matrix and master equation}

As previously noted, the emission rate of scalar quanta by the atoms, given in Eq.~\eqref{emrate}, exhibits a Planckian form, suggesting an underlying thermal nature. In fact, computing the ratio between the emission and absorption rates yields
\begin{equation}
    \frac{R_{\text{em},\mathbf{s}}}{R_{\text{ab},\mathbf{s}}} = e^{-2\pi\omega/\kappa},\label{balance11}
\end{equation}
which matches the form expected for a thermal state satisfying detailed balance, governed by the Boltzmann factor
\begin{equation}
    \frac{R_{\text{em},\mathbf{s}}}{R_{\text{ab},\mathbf{s}}} = e^{-\beta\omega},
\end{equation}
with an effective temperature given by
\begin{equation}
    T= \beta^{-1} = \frac{\kappa}{2\pi} = \beta_D^{-1} = T_D \quad , \quad \kappa = \frac{2}{\alpha}.\label{temp12}
\end{equation}

We thus find that this effective temperature precisely matches the causal diamond temperature $T_D = 1/(\pi\alpha)$ obtained in Eq.~\eqref{sufgrv1}.\\

To better understand the thermal nature of the HBAR, as implied by the results derived within the CQM framework and summarized in Eqs.~\eqref{balance11}–\eqref{temp12}, we consider the system composed of the atomic cloud (A) and the scalar quanta emitted by these atoms, which we denote as the radiation field (RF). The density matrix of the radiation field is obtained by tracing out the atomic degrees of freedom from the total density matrix of the system, i.e., $\hat{\rho}^{\rm RF} = \Tr_A \left[\hat{\rho}^{\rm RF-A}\right]$. For simplicity, we shall henceforth omit the superscript (RF), with the understanding that the analysis focuses exclusively on the radiation field. A remarkable feature is that, for the cloud of freely falling atoms injected at random times, the radiation field density matrix evolves into a diagonal form. 
This leads to a multimode master equation whose structure is general with respect to the background geometry (a detailed discussion can be found in Refs.~\cite{azizi1,azizi2})
\begin{eqnarray}
\dot{\rho}_{\text{diag}}\qty(\qty{n}) = - \sum_j \left[
R_{\text{em},j} \qty[(n_j+1)\rho_{\text{diag}}(\qty{n}) - n_j \rho_{\text{diag}}(\qty{n}_{n_j-1})] \right.\nonumber\\
\left. + R_{\text{ab},j} \qty[n_j \rho_{\text{diag}}(\qty{n}) - (n_j+1) \rho_{\text{diag}}(\qty{n}_{n_j+1})]
\right], \label{fmeq12}
\end{eqnarray}
where the index $j$ labels the field modes $\mathbf{s}_j$. The diagonal density matrix, written in the occupation number basis $\qty{n} \equiv \qty{n_1, n_2, \cdots, n_j, \cdots}$, is given by $\rho_{\text{diag}}(\qty{n}) = \rho_{n_1,n_2,\cdots; \, n_1,n_2,\cdots}$, where $n_j \equiv n_{\mathbf{s}_j}$ is the occupation number of scalar quanta in the mode $\mathbf{s}_j$. The notation $\qty{n}_{n_j + q} \equiv \qty{n_1, n_2, \cdots, n_j + q, \cdots}$ denotes a shift of $q \in \mathbb{Z}$ in the occupation number of the $j$th mode, with all other occupation numbers unchanged.\\

The steady-state density matrix $\rho_{\text{diag}}^{(SS)}(\qty{n})$ is obtained by setting the time derivative in Eq.~\eqref{fmeq12} to zero. To construct this solution, we first consider the single-mode case, for which the steady-state distribution satisfies (see Ref.~\cite{azizi1})
\begin{eqnarray}
    \left. \rho^{(SS)}_{n_j,n_j}\right|_{\text{single-mode}}=\qty[1-\qty(\frac{R_{\text{em},j}}{R_{\text{ab},j}})] \qty(\frac{R_{\text{em},j}}{R_{\text{ab},j}})^{n_j}=\frac{1}{Z_j} e^{-n_j\beta\omega_j},
\end{eqnarray}
where $Z_j = [1 - e^{-\beta\omega_j}]^{-1}$ is the partition function for the single mode $\mathbf{s}_j$, and $\beta = 2\pi/\kappa = \pi\alpha$.\\

Because the atoms are injected randomly and the modes evolve independently, the full multimode steady-state density matrix factorizes as a product over single-mode contributions
\begin{eqnarray}
    \rho^{(SS)}_{\text{diag}}(\{n\})=\prod_j\rho^{(SS)}_{n_j,n_j},
\end{eqnarray}
from which we obtain the explicit thermal form
\begin{eqnarray}
    \rho^{(SS)}_{\text{diag}}(\{n\})=N \prod_j\qty(\frac{R_{\text{em},j}}{R_{\text{ab},j}})^{n_j}=\frac{1}{Z(\beta)} \prod_j e^{-n_j \beta \omega_j},\label{hbarre12}
\end{eqnarray}
where $Z(\beta) = N^{-1} = \prod_j Z_j = \prod_j [1 - e^{-\beta \omega_j}]^{-1}$ is the full partition function.\\

Therefore, the steady-state density matrix of the radiation field corresponds to a thermal distribution at the causal diamond temperature $T_D = 1/(\pi\alpha)$, in agreement with the results of Eqs.~\eqref{balance11}–\eqref{temp12}. It is important to note that this density matrix reproduces the expected Planckian form for the steady-state average occupation number
\begin{equation}
    \expval{n_j}^{(SS)} = \frac{1}{e^{\beta\omega_j} - 1}.
\end{equation}
Importantly, this result depends explicitly on the near-horizon CQM structure, from which the thermal emission and absorption rates emerge. Thus, Eq.~\eqref{hbarre12} extends the validity of the steady-state analysis and the HBAR framework to the CD spacetime geometry.

\section{HBAR thermodynamics}
In this section, we compute the rate of change of the von~Neumann entropy, namely the \emph{entropy flux}, associated with the generation of acceleration radiation (scalar quanta) emitted by the atomic cloud. This corresponds precisely to the HBAR entropy proposed in Ref.~\cite{scully2018quantum}. To proceed, we start with the von Neumann entropy of a quantum system, defined as $S = -\Tr[\rho \ln \rho]$, from which the entropy flux is given by $\dot{S} = -\Tr[\dot{\rho} \ln \rho]$. When applied to the radiation field density matrix, this takes the form
\begin{equation}
    \dot{S} = -\sum_{\{n\}} \dot{\rho}_{\text{diag}}(\qty{n}) \ln\qty[ \rho_{\text{diag}}(\qty{n})].\label{vnebar}
\end{equation}

Near the steady-state configuration, the density matrix inside the logarithm in Eq.~\eqref{vnebar} can be approximated to leading order by $\rho^{(SS)}_{\text{diag}}(\{n\})$ as given in Eq.~\eqref{hbarre12}. In this regime, the entropy flux becomes
\begin{eqnarray}
    \dot{S} &\approx& -\sum_{\{n\}} \dot{\rho}_{\text{diag}}(\qty{n}) \ln\qty[ \rho^{(SS)}_{\text{diag}}(\{n\})]\nonumber\\
    &=& -\sum_{\{n\}} \dot{\rho}_{\text{diag}}(\qty{n}) \ln\qty[ \frac{1}{Z(\beta)} \prod_j e^{-n_j \beta \omega_j}]\nonumber\\
    &=& \sum_j\sum_{\{n\}} \dot{\rho}_{\text{diag}}(\qty{n})\, n_j \beta\omega_j - \sum_j\sum_{\{n\}} \dot{\rho}_{\text{diag}}(\qty{n}) \ln(1 - e^{-\beta\omega_j}),\label{eqdhj1}
\end{eqnarray}
where we used that $Z = \prod_j [1 - e^{-\beta \omega_j}]^{-1}$. Additionally, using the trace normalization condition $\Tr[\rho] = \sum_{\qty{n}} \rho_{\text{diag}}(\qty{n}) = 1$, and the dynamical generalization of the expectation value of the occupation number $\expval{n_j} = \sum_{\qty{n}} n_j \rho_{\text{diag}}(\qty{n})$, Eq.~\eqref{eqdhj1} simplifies to
\begin{eqnarray}
    \dot{S} &=& \sum_j \underbrace{\qty[\sum_{\{n\}} \dot{\rho}_{\text{diag}}(\qty{n})\, n_j]}_{\dot{\expval{n_j}}} \beta \omega_j 
    - \sum_j \underbrace{\qty[\sum_{\{n\}} \dot{\rho}_{\text{diag}}(\qty{n})]}_{=\,0} \ln(1 - e^{-\beta \omega_j})\nonumber\\
    &=& \beta_D \sum_j \dot{\expval{n_j}}\, \omega_j,\label{eqdhj2}
\end{eqnarray}
where we used the fact that the second term vanishes due to constancy of trace normalization, and substituted $\beta \to \beta_D$ as given in Eq.~\eqref{temp12}.\\

The quantity $\dot{\expval{n_j}}\, \omega_j$ represents the contribution to the energy flux carried away by scalar quanta with frequency $\omega_j$ in the acceleration radiation. The total energy flux is thus given by $\dot{E} = \sum_j \dot{\expval{n_j}}\, \omega_j$. Therefore, the HBAR von Neumann entropy flux is
\begin{equation}
    \dot{S} = \beta_D \dot{E},\label{sdotedot1}
\end{equation}
which can be restated in terms of infinitesimal variations as
\begin{equation}
    \delta S = \beta_D \delta E \equiv \delta S^{(\text{th})},\label{enth12}
\end{equation}
which is precisely the thermodynamic entropy variation $\delta S^{(\text{th})}$ associated with the radiation field of scalar quanta. Thus, near equilibrium—where the steady-state approximation provides a good leading-order description—the variation in the HBAR von Neumann entropy coincides with the thermodynamic entropy variation of the radiation field.\\

Finally, we revisit the von Neumann entropy itself. Near the steady-state configuration, the logarithmic term can once again be approximated by $\rho^{(SS)}_{\text{diag}}(\{n\})$. Using the trace normalization condition and the dynamical expectation values as before, we obtain
\begin{eqnarray}
    S &=& \sum_j \left[\sum_{\{n\}} \rho_{\text{diag}}(\{n\})\, n_j\right] \beta \omega_j 
    - \sum_j \left[\sum_{\{n\}} \rho_{\text{diag}}(\{n\})\right] \ln(1 - e^{-\beta\omega_j}) \nonumber\\
    &=& \sum_j \langle n_j \rangle\, \beta \omega_j - \sum_j \ln(1 - e^{-\beta \omega_j}) \nonumber\\
    &=& \beta_D (E - F) \equiv S^{(\text{th})}, \label{entrphif}
\end{eqnarray}
where we have used the definition of the total energy $E = \sum_j \langle n_j \rangle\, \omega_j$, identified $\beta \to \beta_D$, and recognized the function $\beta F = \sum_j \ln(1 - e^{-\beta \omega_j}) = -\ln Z(\beta)$, which corresponds to the Helmholtz free energy. From this, we observe that the von Neumann entropy obtained in Eq.~\eqref{entrphif} is in complete agreement with the flux calculated in Eq.~\eqref{sdotedot1}, since its variation at fixed (causal diamond) temperature $T_D = \beta_D^{-1}$ reduces to the expression given in Eq.~\eqref{enth12}.\\

Therefore, we conclude that the causal horizons of the CD spacetime effectively behave as a \emph{topological thermal reservoir}, in the sense that the temperature $T_D$ characterizing the canonical ensemble of the radiation field arises solely from the global causal structure of spacetime and the corresponding mode decomposition of the quantum field in the Boulware vacuum, rather than from any underlying microscopic degrees of freedom.

\section{Discussion}
In this article, by analyzing the near-horizon region of the CD spacetime, we identified an emergent conformal symmetry governed by CQM, as discussed in Sec.~\ref{seciii}, which allowed for an analytic treatment of the field modes. Besides elucidating the associated entropy dynamics, this symmetry is essential for computing the emission and absorption spectra of scalar quanta [see Eqs.~\eqref{emrate}--\eqref{remrabs12}] produced and absorbed by the freely falling atomic cloud as it approaches the causal horizon of the CD spacetime. Employing tools from quantum optics, we computed the von Neumann entropy of the radiation field—composed of scalar quanta—and its corresponding flux, given in Eq.~\eqref{sdotedot1}.\\

We showed that the HBAR entropy flux satisfies the thermodynamic relation $\delta S = \beta_D \delta E$, and that the von Neumann entropy reproduces the canonical expression $S = \beta_D (E - F)$, where $F$ denotes the Helmholtz free energy. These results demonstrate that the causal horizons of the CD spacetime effectively behave
as a topological thermal reservoir at temperature $T_D = \kappa/(2\pi)$, with $\kappa$ being the effective surface gravity defined in Eq.~\eqref{sufgrv1}. In this way, the entropy associated with HBAR in this context coincides with the thermodynamic entropy of a canonical ensemble, thereby providing a consistent thermodynamic interpretation of acceleration radiation in the CD spacetime.\\

Finally, our findings highlight that the essential features of HBAR—originally formulated for black hole geometries—extend naturally to the CD spacetime, revealing that the presence of causal horizons alone suffices to produce thermal radiation and entropy flux. This highlights that HBAR is fundamentally a manifestation of causal structure rather than being exclusive to black holes, and opens new avenues for exploring quantum thermodynamics, information flow, and entropy production in finite-lifetime spacetimes with causal horizons.

\begin{acknowledgments}
We would like to thank the attendees of the 2025 TAMU-Princeton-Caspar-Baylor-CSU-UIUC Summer School, where these and other results were first presented, for their interest and many questions. We are especially grateful to Wolfgang Schleich for his careful reading of a draft version of the paper, for his thoughtful questions and suggestions and for his encouragement to publish these results. All the authors of this paper were partially supported by the Army Research Office (ARO) under Grant No. W911NF-23-1-0202. In addition, G.V.-M. gratefully acknowledges the Center for Mexican American and Latino/a Studies at the University of Houston for their generous support through a Lydia Mendoza Fellowship.
\end{acknowledgments}

\appendix
\section{EMISSION RATE FOR AN ATOMIC CLOUD APPROACHING $\mathcal{H}^+_\infty$}\label{ap12}
For completeness, we now compute the emission and absorption ratios corresponding to the process in which atoms are randomly injected from the asymptotic past, $\eta \to -\infty$, at $\rho \ll 1$, directed toward $\mathcal{H}^+_\infty$, i.e., $\rho \gg 1$. \\

Near the horizon at $\rho \gg 1$, the field equation reduces to its radial component given in Eq.~\eqref{cqmeqrho0}, which is precisely the CQM equation. Therefore, the solution coincides with that obtained for $\rho \ll 1$, given in Eq.~\eqref{cqmeom11}, namely
\begin{equation}
\frac{\psi(\rho)}{\sqrt{\rho}} \sim \rho^{\pm i\Theta},
\end{equation}
where $(+)$ corresponds to ingoing modes, while $(-)$ corresponds to outgoing modes. This contrasts with the case in which the atoms are directed toward the horizon $\mathcal{H}^+_0$. \\

Having established the CQM modes of the field, we can now compute the corresponding emission rate (and, consequently, the absorption rate). To this end, we need to characterize $\eta$ and $\tau$ in terms of $\rho$ in the region $\rho \gg 1$. In our discussion of the geometry in Sec.~\ref{geonhap1}, we found that by performing the substitution $4\tilde{\rho} = \alpha^2 / \rho$, the near-horizon analysis in the region of interest is simplified in terms of $\tilde{\rho} \ll 1$ [see Eq.~\eqref{etanadkambdar1}]. In this parametrization, $\eta$, $\tau$, and the CQM field modes take the form
\begin{eqnarray}
\eta &=& -\frac{1}{\kappa} \ln \tilde{\rho} + \mathcal{O}(\tilde{\rho}^2) + \text{const.},\\ 
\tau &=& -\frac{8 \kappa}{\tilde{e}} \tilde{\rho}^2 + \mathcal{O}(\tilde{\rho}^4) + \text{const.},\\
\phi_\omega^{\pm\,(\mathrm{CQM})}(\eta,\tilde{\rho}) &\sim& e^{-i\omega\eta}\tilde{\rho}^{\mp i\Theta},
\end{eqnarray}
where, in terms of $\tilde{\rho}$, $(+)$ now corresponds to outgoing modes, while $(-)$ corresponds to ingoing modes. One can thus recognize the complete analogy with the case $\rho \ll 1$. \\

As discussed earlier in the article, only outgoing modes contribute to the emission process, namely those corresponding to scalar quanta emitted by the atoms and propagating away from the causal horizon $\mathcal{H}^+_\infty$ toward $\rho \ll 1$. The emission rate is therefore given by
\begin{eqnarray}
\int d\tau\, [\phi^{+(\text{CQM})}_{\omega}(\eta, \tilde{\rho})]^* e^{i\nu\tau} 
&=& \int d\tau\, e^{i\omega\eta} \tilde{\rho}^{-i\Theta} e^{i\nu\tau} \nonumber\\
&=& \frac{16\kappa}{\tilde{e}} \int_0^{\tilde{\rho}_\circ} d\rho\, \tilde{\rho}\,\tilde{\rho}^{-2i\omega/\kappa} e^{- i8\nu\kappa\tilde{\rho}^2/\tilde{e}},
\end{eqnarray}
where $\tilde{\rho}_\circ$ enforces the constraint on $\tilde{\rho}$. Following the same reasoning as in Sec.~\ref{emabprb1}, the upper limit may be extended, allowing the integral to be evaluated and the emission rate obtained.\\

Finally, we arrive at
\begin{equation}
R_{\text{em},\mathbf{s}} = \frac{2\pi \mathfrak{r} g^2 \omega}{\kappa \nu^2} \left( \frac{1}{e^{2\pi\omega/\kappa} - 1} \right).
\end{equation}
Thus, the result is identical for the two cases: when the atom approaches the horizon $\mathcal{H}^+_0$, and when it approaches $\mathcal{H}^+_\infty$.

\bibliography{apssamp}

\end{document}